\documentclass[12pt,a4paper]{article}
\usepackage{amsfonts}
\usepackage{epsfig,amssymb,euscript}
\usepackage{amsmath,amscd}
\usepackage{appendix}

\numberwithin{equation}{section}
\newcommand{\be}{\begin{eqnarray}}
\newcommand{\ee}{\end{eqnarray}}
\newcommand{\bea}{\begin{eqnarray}}
\newcommand{\eea}{\end{eqnarray}}
\newcommand{\ba}{\begin{array}}
\newcommand{\ea}{\end{array}}
\newcommand{\nn}{\nonumber \\}

\newcommand{\hook}{{\setlength{\unitlength}{11pt}   % adjust pt size here
                    \begin{picture}(.833,.8)
                    \put(.15,.08){\line(1,0){.35}}
                    \put(.5,.08){\line(0,1){.5}}
                    \end{picture}}}

\def\bo{{\bar{1}}}
\def\bt{{\bar{2}}}

\begin{document}

\begin{titlepage}

\vfill

\begin{flushright}
%Imperial/TP/2008/JG/02\
\end{flushright}

\vfill

\begin{center}
   \baselineskip=16pt
  {\Large\bf Cosmological Einstein-Maxwell Instantons and Euclidean Supersymmetry: Beyond Self-Duality}
   \vskip 2cm
      M.  Dunajski$^{1}$, J. B.  Gutowski$^2$, W. A. Sabra$^3$
      and Paul Tod$^4$\\
 \vskip .6cm
  \begin{small}
   $^1$\textit{Department of Applied Mathematics and Theoretical Physics
 \\
        University of Cambridge
 \\
        Wilberforce Road, Cambridge, CB3 0WA, UK \\
        E-mail: m.dunajski@damtp.cam.ac.uk}
        \end{small}\\*[.6cm]
      \begin{small}
      $^2$\textit{  Department of Mathematics, King' s College London
 \\
         Strand, London WC2R 2LS, UK.\\
     E-mail: jan.gutowski@kcl.ac.uk}
        \end{small}\\*[.6cm]
      \begin{small}
      $^3$\textit{Centre for Advanced Mathematical Sciences and
        Physics Department, \\
        American University of Beirut, Lebanon \\
        E-mail: ws00@aub.edu.lb}
         \end{small}\\*[.6cm]
\begin{small}
$^4$\textit{ The Mathematical Institute, Oxford University
 \\
         24-29 St Giles, Oxford OX1 3LB, UK.\\
     E-mail: paul.tod@sjc.ox.ac.uk}
        \end{small}
\end{center}
\vfill

\begin{center}
\textbf{Abstract}
\end{center}

\begin{quote}
 We construct new supersymmetric solutions to the Euclidean 
Einstein-Maxwell theory with a non-vanishing cosmological constant, and for which the
Maxwell field strength is neither self-dual or anti-self-dual.
We find that there are three classes of solutions, depending on the sign of the Maxwell field strength 
and cosmological constant terms in the Einstein equations which arise from the integrability
conditions of the Killing spinor equation.
The first class is a Euclidean version of a Lorentzian supersymmetric
solution found in \cite{lorentz1, lorentz2}.
The second class is constructed from a three dimensional base space
which admits a hyper-CR Einstein-Weyl structure. 
The third class  is the Euclidean Kastor-Traschen solution.

\end{quote}

\vfill

\end{titlepage}

%%%%%%%%%%%%%%%%%%%%%%%%%%%%%%%%%%%%%%%%%%%%%%%%%%%%%%%%%%%%%%%%%%%%%%%%%%

%%%%%%%%%%%%%%%%%%%%%%%%%%%%%%%%%%%%%%%%%%%%%%%%%%%%%%%%%%%%%%%%%%%%%%%%%%

\section{Introduction}

A considerable amount of work has been done in recent years towards the
classification of solutions admitting supersymmetry in various
supergravity theories. The classification of supersymmetric solutions was
initiated in the work of \cite{ghgr, Tod83}. There, Einstein-Maxwell theory was
considered as the bosonic sector of $N=2$ supergravity in four dimensions.
In \cite{Tod83} solutions with time-like and null Killing vectors, admitting
a supercovariantly constant spinor were determined.

More recently there has been some work on the classification of
solutions of Euclidean Einstein-Maxwell theory in the
case with a zero \cite{DH07, instantons}, and non-zero \cite{firstpaper}
cosmological constant. The motivation for studying such Euclidean solutions is their
possible relevance to the non-perturbative analysis in the theory of quantum
gravity. In the early studies of gravitational instantons, the Einstein
equations of motion, which are in general hard to solve, were simplified by
assuming a self-dual Riemann tensor. This is analogous to the condition of
self-dual Yang-Mills field strengths imposed in the study of instantons in 
\cite{polyakov}.

The analysis of \cite{firstpaper} was mainly focused on the cases where the
Maxwell field is anti-self-dual, where it was shown that the
field equations of supersymmetric solutions reduce to a sub-class of Einstein-Weyl
equations in three dimensions \cite{mtconf} which is integrable by 
twistor transform \cite{Hi,JT85}. For supersymmetric Euclidean solutions, it turns out that the
anti-selfduality of the Maxwell field implies the conformal
anti-selfduality (ASD) of the Weyl tensor. The Killing spinor equations
used in the analysis of \cite{firstpaper} contain a continuous parameter.
For a particular value of this parameter, the solutions constructed have a
Killing vector and are related to solutions of the $SU(\infty )$ Toda
equation. 

In our present work, the anti-self-duality condition on the Maxwell
field is relaxed and we classify solutions
admitting Killing spinors using spinorial geometry techniques which were 
first used  to analyse supersymmetric solutions in ten and eleven dimensions in
\cite{first},  partly based on \cite{lwh}.
Other applications of spinorial geometry techniques are
the classification of solutions in lower dimensions 
\cite{recentlower}; the first systematic classification of supersymmetric
extremal black hole near-horizon geometries in ten-dimensional heterotic
supergravity \cite{hethor}; and the classification of supersymmetric
solutions of Euclidean $N=4$ \ super Yang-Mills theory \cite{dietmar1}.

We plan our work as follows. In section two, we write down a Killing spinor
equation for the Euclidean Einstein-Maxwell theory with a cosmological
constant. The Killing spinor equation is fixed by considering the associated
integrability condition and comparing with the Einstein equations of motion.
Four possibilities for the Killing spinor equation are obtained, corresponding
to the different possible choices for signs of the Maxwell and cosmological constant
terms in the Einstein field equations.
In section
two we also write down the essential equations needed for our analysis based
on spinorial geometry. In sections three and four we present the
solutions obtained through the analysis of the Killing spinor equation.
The solutions in section three are a Euclidean version
of the ``timelike" class of solutions found in \cite{lorentz1, lorentz2}.
The solutions in section four consist of the Euclidean Kastor-Traschen solution,
and a set of solutions constructed from a 3-dimensional base space which admits
a hyper-CR Einstein-Weyl structure. In section 5 we present our conclusions.

\section{The Killing Spinor Equations}

Let $g$ be a positive definite metric on a Riemannian four manifold, and
let $\Gamma_\mu$ be the generators of the Clifford algebra, $\{\Gamma_\mu, \Gamma_\nu \}=2 \delta_{\mu \nu}$.
Unless stated otherwise, we shall work in an orthonormal frame, with frame indices $\mu, \nu$ in which
$g_{\mu \nu}=\delta_{\mu \nu}$. To begin, we consider a generalized Killing spinor equation: 
\begin{equation}
\nabla _{\mu }\epsilon =\left( c_{1}+d_{1}\gamma _{5}\right) F_{\nu _{1}\nu
_{2}}\Gamma ^{\nu _{1}\nu _{2}}\Gamma _{\mu }\epsilon +(c_{2}+d_{2}\gamma
_{5})\Gamma _{\mu }\epsilon +(c_{3}+d_{3}\gamma _{5})A_{\mu }\epsilon
\label{gkse}
\end{equation}
for complex constants $c_{1},c_{2},c_{3},d_{1},d_{2},d_{3}$, and $\Gamma_{\mu \nu}=\Gamma_{[\mu} \Gamma_{\nu]}$.
In addition,
$\nabla _{\mu}\epsilon  \equiv (\partial _{\mu }+{\frac{1}{4}}\Omega _{\mu ,\nu _{1}\nu
_{2}}\Gamma ^{\nu _{1}\nu _{2}})\epsilon $ is the supercovariant derivative
with spin connection $\Omega $, and $F=dA$ is the gauge field strength
satisfying
\begin{equation}
dF=0,\qquad d\star F=0.
\end{equation}
We proceed to evaluate the integrability conditions associated with 
({\ref{gkse}}). By examining the terms which are linear in the gauge potential $A$%
, one must have 
\begin{equation}
d_{3}=0.
\end{equation}
The integrability conditions further imply that 
\begin{equation}
R_{\mu \nu }+32(c_{1}^{2}-d_{1}^{2})F_{\mu \sigma }F_{\nu }{}^{\sigma
}+\left( 12(c_{2}^{2}-d_{2}^{2})-8(c_{1}^{2}-d_{1}^{2})F_{\nu _{1}\nu
_{2}}F^{\nu _{1}\nu _{2}}\right) g_{\mu \nu }=0,  \label{intx}
\end{equation}
with $c_{3}$ fixed by 
\begin{equation}
c_{3}=8(c_{1}c_{2}-d_{1}d_{2}),
\end{equation}
and 
\begin{equation}
c_{1}d_{2}-d_{1}c_{2}=0.  \label{van2}
\end{equation}
Next, consider a re-definition of the spinor via 
\begin{equation}
\epsilon =(a+b\gamma _{5})\epsilon ^{\prime }
\end{equation}
for complex constants $a,b$, such that $a^{2}-b^{2}\neq 0$. Substitute this
spinor into ({\ref{gkse}}) and multiply on the left by $(a+b\gamma
_{5})^{(-1)}$ to obtain 
\begin{equation}
\nabla _{\mu }\epsilon ^{\prime }=(c_{1}^{\prime }+d_{1}^{\prime }\gamma
_{5})F_{\nu _{1}\nu _{2}}\Gamma ^{\nu _{1}\nu _{2}}\Gamma _{\mu }\epsilon
^{\prime }+(c_{2}^{\prime }+d_{2}^{\prime }\gamma _{5})\Gamma _{\mu
}\epsilon^{\prime} +8(c_{1}c_{2}-d_{1}d_{2})A_{\mu }\epsilon^{\prime}  \label{gkse2}
\end{equation}
where 
\begin{eqnarray}
c_{1}^{\prime } &=&{\frac{1}{(a^{2}-b^{2})}}((a^{2}+b^{2})c_{1}-2abd_{1}),
\qquad d_{1}^{\prime }={\frac{1}{(a^{2}-b^{2})}}
(-2abc_{1}+(a^{2}+b^{2})d_{1}),\newline
\notag \\
c_{2}^{\prime } &=&{\frac{1}{(a^{2}-b^{2})}}((a^{2}+b^{2})c_{2}-2abd_{2}),
\qquad d_{2}^{\prime }={\frac{1}{(a^{2}-b^{2})}}
(-2abc_{2}+(a^{2}+b^{2})d_{2})\newline
\text{ }.  \notag \\
&&
\end{eqnarray}
Note in particular that 
\begin{eqnarray}
(c_{1}^{\prime })^{2}-(d_{1}^{\prime })^{2}
&=&(c_{1})^{2}-(d_{1})^{2},\qquad (c_{2}^{\prime })^{2}-(d_{2}^{\prime
})^{2}=(c_{2})^{2}-(d_{2})^{2}\newline
\notag \\
c_{1}^{\prime }c_{2}^{\prime }-d_{1}^{\prime }d_{2}^{\prime }
&=&c_{1}c_{2}-d_{1}d_{2},\qquad \ \ \ c_{1}^{\prime }d_{2}^{\prime
}-d_{1}^{\prime }c_{2}^{\prime }=c_{1}d_{2}-d_{1}c_{2}\text{ \ }.
\end{eqnarray}
In order for the integrability condition ({\ref{intx}}) to correspond to the
Einstein-Maxwell equations 
\be
\label{emax1}
R_{\mu \nu} + 6 \Lambda g_{\mu \nu} + c \big( 4 F_{\mu \sigma} F_{\nu}{}^\sigma - g_{\mu \nu}
F_{\nu_1 \nu_2} F^{\nu_1 \nu_2} \big) =0
\ee
for non-zero constants $\Lambda, c$,
one must
have $c_{1}\neq \pm d_{1}$. Hence, it is straightforward to show that,
without loss of generality, one can choose $a,b$ with $a\neq \pm b$ such
that $d_{1}^{\prime }=0$. On dropping the primes throughout, one then
obtains $d_{2}=0$ from ({\ref{van2}}). After making this transformation, the
Killing spinor equation becomes: 
\begin{equation}
\nabla _{\mu }\epsilon =c_{1}F_{\nu _{1}\nu _{2}}\Gamma ^{\nu _{1}\nu
_{2}}\Gamma _{\mu }\epsilon +c_{2}\Gamma _{\mu }\epsilon +8c_{1}c_{2}A_{\mu
}\epsilon \ . \label{gkse2b}
\end{equation}
We shall call backgrounds which admit a spinor $\epsilon$ satisfying this equation
supersymmetric. The Killing spinor equation ({\ref{gkse2b}}) has the associated integrability condition 
\begin{equation}
R_{\mu \nu }+32c_{1}^{2}F_{\mu \sigma }F_{\nu }{}^{\sigma }+
\big(12c_{2}^{2}-8c_{1}^{2}F_{\nu _{1}\nu _{2}}F^{\nu _{1}\nu _{2}}\big)g_{\mu
\nu }=0.  \label{intx2}
\end{equation}
Observe that $c_{1},c_{2}$ are then fixed, up to a sign, by comparison with
the Einstein-Maxwell equations ({\ref{emax1}}). We consider four cases, corresponding to 
$(c_{1},c_{2})=(-{\frac{i}{4}},-{\frac{1}{2\ell }})$, $(c_{1},c_{2})=(
-{\frac{i}{4}},-{\frac{i}{2\ell }})$, $(c_{1},c_{2})=(-{\frac{1}{4}},
-{\frac{1}{2\ell }})$ and $(c_{1},c_{2})=(-{\frac{1}{4}},-{\frac{i}{2\ell }})$ for 
$\ell \in \mathbb{R}$. Note that the cosmological constant $\Lambda$
appearing in the Einstein-Maxwell equations ({\ref{emax1}}) is given by $\Lambda = 2 c_2^2$, and so
$\ell^{-2} = \pm 2 \Lambda$, depending on whether $c_2$ is real or imaginary. 
We remark that in \cite{firstpaper} a similar analysis was carried out
in the case for which $F$ is anti-self-dual. It was shown that when
this restriction is made, the conditions imposed  on the constants appearing in
the Killing spinor equation, which one obtains by comparing the integrability
conditions with the Einstein-Maxwell equations, are weaker. In particular, in addition to the cosmological
constant, there is an extra free real parameter in the Killing spinor equations.
However, when one drops the anti-self-dual condition on $F$,
the conditions obtained from matching the integrability conditions to the Einstein-Maxwell equations
are stronger, and as we have shown, the only free parameter remaining, after re-scaling of $F$
is taken into account, is the cosmological constant.

In order to analyse the solutions of the Killing spinor equation 
({\ref{gkse2}}), it will be convenient to work with a Hermitian basis $\mathbf{e}^{1},\mathbf{e}^{2},\mathbf{e}^{\bar{1}},\mathbf{e}^{\bar{2}}$, with ${\mathbf{e}}^{\bar{1}} = {\overline{\mathbf{e}^1}}, {\mathbf{e}}^{\bar{2}} = {\overline{\mathbf{e}^2}}$,
for which the spacetime metric is\bigskip
\begin{equation}
ds^{2}=2\left( \mathbf{e}^{1}\mathbf{e}^{\bar{1}}+\mathbf{e}^{2}
\mathbf{e}^{\bar{2}}\right) .
\end{equation}
The metric has signature $(+,+,+,+)$. The space of Dirac spinors is the
complexified space of forms on $\mathbb{R}^{2}$, with basis 
$\{1,e_{1},e_{2},e_{12}=e_{1}\wedge e_{2}\}$; a generic Dirac spinor 
$\epsilon $ is a complex linear combination of these basis elements. In this
basis, the action of the Gamma matrices on the Dirac spinors is given by 
\begin{equation}
\Gamma _{m}=\sqrt{2}{e_{m}}\hook ,\qquad \Gamma _{\bar{m}}=\sqrt{2}e_{m}\wedge
\end{equation}
for $m=1,2$, where $\hook$ denotes contraction. We define 
\begin{equation}
\gamma _{5}=\Gamma _{1\bar{1}2\bar{2}}
\end{equation}
which acts on spinors via 
\begin{equation}
\gamma _{5}1=1,\qquad \gamma _{5}e_{12}=e_{12},\qquad \gamma
_{5}e_{m}=-e_{m}\quad m=1,2.
\end{equation}
To proceed we examine the orbits of $Spin(4)=Sp(1)\times Sp(1)$ acting on
the space of Dirac spinors. In particular, as observed (for example) in 
\cite{bryant}, there are three non-trivial orbits. In our notation, following the
reasoning set out in \cite{half2007}, one can use $SU(2)$ transformations to
rotate a generic spinor $\epsilon $ into the canonical form 
\begin{equation}
\epsilon =\lambda +\sigma e_{1}  \label{ksp}
\end{equation}
where $\lambda ,\sigma \in \mathbb{R}$. The three orbits 
correspond to the cases $\lambda =0$, $\sigma \neq 0$; $\lambda \neq 0$, 
$\sigma =0$ and $\lambda \neq 0$, $\sigma \neq 0$.

\section{Solutions with $c_1=-{\frac{i }{4}}, c_2=-{\frac{1 }{2 \ell}}$}

In this case, the Killing spinor equation is 
\begin{equation}
\left( \partial _{\mu }+{\frac{1}{4}}\Omega _{\mu ,\nu _{1}\nu _{2}}\Gamma
^{\nu _{1}\nu _{2}}+{\frac{i}{4}}F_{\nu _{1}\nu _{2}}\Gamma ^{\nu _{1}\nu
_{2}}\Gamma _{\mu }+{\frac{1}{2\ell }}\Gamma _{\mu }-{\frac{i}{\ell }}A_{\mu
}\right) \epsilon =0.  \label{kse}
\end{equation}
To proceed, one evaluates the Killing spinor equation ({\ref{kse}}) acting
on the spinor ({\ref{ksp}}). First we consider the special cases for which 
$\epsilon =\lambda 1$ ($\lambda \neq 0$) or $\epsilon =\sigma e_{1}$ ($\sigma
\neq 0$). In the former case, on substituting $\sigma =0$ into the Killing
spinor equation, we obtain the following conditions 
\begin{eqnarray}
\lambda \left( {\frac{i}{\sqrt{2}}}(F_{1\bar{1}}-F_{2\bar{2}})+
{\frac{1}{\sqrt{2}}}\ell ^{-1}\right) &=&0,  \notag \\
\lambda \left( -{\frac{i}{\sqrt{2}}}(F_{1\bar{1}}-F_{2\bar{2}})+
{\frac{1}{\sqrt{2}}}\ell ^{-1}\right) &=&0,
\end{eqnarray}
which admit no solution. In the latter case, on substituting $\lambda =0$,
we obtain 
\begin{eqnarray}
\sigma \left( -{\frac{i}{\sqrt{2}}}(F_{1\bar{1}}+F_{2\bar{2}})+{\frac{1}{\sqrt{2}}}\ell
^{-1}\right) &=&0,  \notag \\
\sigma \left( -{\frac{i}{\sqrt{2}}}(F_{1\bar{1}}+F_{2\bar{2}})-{\frac{1}{\sqrt{2}}}\ell
^{-1}\right) &=&0,
\end{eqnarray}
which again admit no solution. It follows that there are no
supersymmetric solutions corresponding to these cases. Note that this
analysis has not made use of any reality conditions, and hence these types
of solutions are excluded for all choices of $(c_{1},c_{2})$. It remains to
analyse ({\ref{kse}}) in the case for which $\lambda \neq 0$ and $\sigma \neq
0$. In this case, one obtains the following geometric conditions: 
\begin{eqnarray}
2\partial _{1}\lambda -\lambda \Omega _{2,\bar{1}\bar{2}}+\sqrt{2}\ell
^{-1}\sigma &=&0,  \notag \\
2\partial _{1}\sigma +\sigma \Omega _{2,\bar{2}1}+\sqrt{2}\ell ^{-1}\lambda
&=&0,  \notag \\
2\partial _{2}\lambda +\lambda \Omega _{1,12} &=&0,  \notag \\
2\partial _{2}\sigma -\sigma \Omega _{\bar{1}2\bar{1}} &=&0,  \notag \\
\Omega _{1,2\bar{1}}=\Omega _{1,\bar{1}\bar{2}}=\Omega _{2,12}=
\Omega _{2,2\bar{1}} &=&0,  \notag \\
2\Omega _{2,1\bar{1}}+\Omega _{\bar{1},2\bar{1}}+\Omega _{1,12} &=&0,  \notag
\\
\lambda \sigma (-2\Omega _{1,1\bar{1}}-\Omega _{2,\bar{1}\bar{2}}+
\Omega _{2,\bar{2}1})+\sqrt{2}\ell ^{-1}(\lambda ^{2}+\sigma ^{2}) &=&0,  \label{geo1}
\end{eqnarray}
the following conditions on the gauge potential 
\begin{eqnarray}
A_{1} &=&{\frac{i\ell }{2}}\left( \Omega _{1,2\bar{2}}+{\frac{1}{2}}\Omega
_{2,\bar{1}\bar{2}}+{\frac{1}{2}}\Omega _{2,\bar{2}1}+{\frac{1}{\sqrt{2}\ell
\lambda \sigma }}(\lambda ^{2}-\sigma ^{2})\right) ,  \notag \\
A_{2} &=&{\frac{i\ell }{2}}\left( \Omega _{2,2\bar{2}}+{\frac{1}{2}}\Omega
_{1,12}-{\frac{1}{2}}\Omega _{\bar{1},2\bar{1}}\right) ,  \label{gpot1}
\end{eqnarray}
and on the gauge field strength 
\begin{eqnarray}
\lambda \Omega _{1,12}+\sqrt{2}i\sigma F_{12} &=&0,  \notag \\
\sigma \Omega _{1,\bar{2}1}+\sqrt{2}i\lambda F_{1\bar{2}} &=&0,  \notag \\
\sigma \Omega _{2,\bar{2}1}+\frac{\lambda }{\sqrt{2}}\left( -{i}(F_{1\bar{1}}
-F_{2\bar{2}})+{\frac{1}{\ell }}\right) &=&0,  \notag \\
\lambda \Omega _{2,\bar{1}\bar{2}}-\frac{\sigma }{\sqrt{2}}
\left( {i}F_{q}{}^{q}+{\frac{1}{\ell }}\right) &=&0.  \label{gfs}
\end{eqnarray}
To proceed with the analysis, observe that ({\ref{geo1}}) and ({\ref{gfs}})
imply that 
\begin{equation}
W=i\lambda \sigma (\mathbf{e}^{1}-\mathbf{e}^{\bar{1}})
\end{equation}
defines a Killing vector. Furthermore, one can without loss of generality
make a $U(1)$ transformation (combined with appropriately chosen $SU(2)$
gauge transformations which leave the Killing spinor invariant), and take 
\begin{equation}
W \hook A={\frac{1}{\sqrt{2}}}(\lambda ^{2}-\sigma ^{2}).  \label{gauge1}
\end{equation}
It is then straightforward to show that ({\ref{geo1}}) implies that 
\begin{equation}
\mathcal{L}_{W}\mathbf{e}^{1}=\mathcal{L}_{W}\mathbf{e}^{2}=0,
\end{equation}
and furthermore 
\begin{equation}
\mathcal{L}_{W}\lambda =\mathcal{L}_{W}\sigma =0.
\end{equation}
Observe also that ({\ref{geo1}}) implies that 
\begin{equation}
d\left( \lambda \sigma (\mathbf{e}^{1}+\mathbf{e}^{\bar{1}})\right) =0,
\label{xc}
\end{equation}
and 
\begin{equation}
d(\lambda \sigma \mathbf{e}^{2})=-\ell ^{-1}\left( {2i}A+\sqrt{2}
({\frac{\lambda }{\sigma }}\mathbf{e}^{1}+{\frac{\sigma }{\lambda }}
\mathbf{e}^{\bar{1}})\right) \wedge (\lambda \sigma \mathbf{e}^{2}).  \label{e2der}
\end{equation}
Further simplification can be obtained by noting that one
can apply a $U(1)\times SU(2)$ transformation $e^{-i\Theta }e^{i\Theta
\Gamma _{2\bar{2}}}$ for $\Theta \in \mathbb{R}$ such that
 $\mathcal{L}_{W}\Theta =0$ and work in a gauge for which 
\begin{equation}
A_{1}+A_{\bar{1}}=0
\end{equation}
while preserving the gauge condition ({\ref{gauge1}}). So, in this gauge 
\begin{equation}
A_{1}={\frac{i}{2\sqrt{2}\lambda \sigma }}(\lambda ^{2}-\sigma ^{2})
\label{g1}
\end{equation}
and on substituting this into ({\ref{gpot1}}) one obtains the following
extra constraint on the spin connection 
\begin{equation}
\Omega _{1,2\bar{2}}+{\frac{1}{2}}\Omega _{2,\bar{1}\bar{2}}+{\frac{1}{2}}
\Omega _{2,\bar{2}1}=0.  \label{extra1}
\end{equation}
Next we introduce co-ordinates, and re-write all the conditions in these
co-ordinates. First, introduce a real local co-ordinate $\psi $ such that 
\begin{equation}
W={\frac{\partial }{\partial \psi }}\text{ .}
\end{equation}
Note that all components of the spin connection, gauge potential and 
$\lambda $, $\sigma $ are independent of $\psi $. Note that ({\ref{xc}})
implies there is a real co-ordinate $x$ such that 
\begin{equation}
\lambda \sigma (\mathbf{e}^{1}+\mathbf{e}^{\bar{1}})=dx
\end{equation}
and ({\ref{e2der}}) implies that 
\begin{equation}
\lambda \sigma \mathbf{e}^{2}=Hdz
\end{equation}
for complex $H,z$. In fact, working in the gauge given by ({\ref{g1}}), it
is straightforward to see that ({\ref{e2der}}) implies that, without loss of
generality, one can take $H\in \mathbb{R}$. Furthermore, we write

\begin{equation}
\mathbf{e}^{1}-\mathbf{e}^{\bar{1}}=-2i\lambda \sigma (d\psi +\phi )
\end{equation}
where $\phi =\phi _{x}dx+\phi _{z}dz+\phi _{\bar{z}}d\bar{z}$ is a real
1-form. The functions $\lambda $, $\sigma $, $H$ and the 1-form $\phi $ are
all independent of $\psi $. To proceed, note that the geometric conditions 
({\ref{geo1}}) and ({\ref{extra1}}) are equivalent to: 
\begin{equation}
\partial _{x}\log H=-{\frac{1}{\sqrt{2}\ell }}(\lambda ^{-2}+\sigma ^{-2})
\label{cc1}
\end{equation}
and 
\begin{eqnarray}
d\phi =-{\frac{i}{(\lambda \sigma )^{2}}}(\partial _{z}\log \frac{\lambda 
}{\sigma })dx\wedge dz+{\frac{i}{(\lambda \sigma )^{2}}}(\partial _{\bar{z}}
\log \frac{\lambda }{\sigma })dx\wedge d\bar{z}
\notag \\
+\left( \frac{2iH^{2}}
{(\lambda \sigma )^{2}}\partial _{x}\log \frac{\sigma }{\lambda }\right)
dz\wedge d\bar{z}  +i\sqrt{2}\ell ^{-1}\frac{H^{2}}{\left( \lambda \sigma \right) ^{2}}\left( 
\frac{{1}}{\sigma ^{2}}-\frac{{1}}{\lambda ^{2}}\right) dz\wedge d\bar{z}.
\label{cc2}
\end{eqnarray}
The integrability condition associated with ({\ref{cc2}}) is given by

\begin{eqnarray}
&&\partial _{x}\left( {\frac{H^{2}}{(\lambda \sigma )^{3}}}\left( -2\sigma
\partial _{x}\lambda +2\lambda \partial _{x}\sigma +\sqrt{2}\ell ^{-1}\left( 
{\frac{\lambda }{\sigma }}-{\frac{\sigma }{\lambda }}\right) \right) \right)
\notag \\
&&-{\frac{2}{(\lambda \sigma )^{3}}}(\sigma \partial _{z}\partial _{\bar{z}}
\lambda -\lambda \partial _{z}\partial _{\bar{z}}\sigma )+
{\frac{6}{(\lambda \sigma )^{4}}}(\sigma ^{2}\partial _{z}\lambda \partial _{\bar{z}}
\lambda -\lambda ^{2}\partial _{z}\sigma \partial _{\bar{z}}\sigma )  =0 \ .
\label{cc2a}
\end{eqnarray}
Next, consider the gauge potential $A$ fixed by ({\ref{g1}}) and 
({\ref{gpot1}}); we find 
\begin{equation}
A={\frac{1}{\sqrt{2}}}(\lambda ^{2}-\sigma ^{2})(d\psi +\phi )-{\frac{i\ell 
}{2}}\left( \partial _{z}\log Hdz-\partial _{\bar{z}}\log Hd\bar{z}\right) .
\end{equation}
On comparing $dA$ with $F$ given in ({\ref{gfs}}), we find that the previous
constraints imply that all components of $dA$ agree with $F$ with no further
constraint, with the exception of the $2\bar{2}$ component, from which we
find 
\begin{equation}
H^{-2}\partial _{z}\partial _{\bar{z}}\log H=\frac{\sqrt{2}}{2}\ell
^{-1}\partial _{x}\left( \frac{1}{\lambda ^{2}}+\frac{1}{\sigma ^{2}}\right)
-{\frac{1}{\ell ^{2}\lambda ^{2}\sigma ^{2}}}\left( \left( {\frac{\lambda }{\sigma }}
\right) ^{2}+\left( {\frac{\sigma }{\lambda }}\right) ^{2}-1\right)
.  \label{cc3}
\end{equation}
Finally, it remains to compute the gauge field equations. Note that it is
most straightforward to impose these by requiring that $F-\star F$ is
closed. From this condition one obtains the final constraint 
\begin{equation}
\left( \partial _{z}\partial _{\bar{z}}+H^{2}\partial _{x}\partial
_{x}\right) \lambda ^{-2}-{\frac{3\sqrt{2}H^{2}}{\ell }}\lambda
^{-2}\partial _{x}\left( \lambda ^{-2}\right) +{\frac{2H^{2}}{\ell
^{2}\lambda ^{6}}}=0.  \label{cc4}
\end{equation}
To summarize, on setting $H=e^u$, one finds that
the metric and gauge potential are given by 
\begin{eqnarray}
ds^{2} &=&2\lambda ^{2}\sigma ^{2}(d\psi +\phi )^{2}+{\frac{1}{\lambda
^{2}\sigma ^{2}}}\left( {\frac{1}{2}}dx^{2}+2 e^{2u} dzd\bar{z}\right)  \notag
\\
A &=&{\frac{1}{\sqrt{2}}}(\lambda ^{2}-\sigma ^{2})(d\psi +\phi )-
{\frac{i\ell }{2}}\partial _{z} u dz+{\frac{i\ell }{2}}\partial _{\bar{z}} u d\bar{z} 
 \label{fmm1}
\end{eqnarray}
where $\lambda ,\sigma ,u$ are functions, and $\phi =\phi _{x}dx+\phi
_{z}dz+\phi _{\bar{z}}d\bar{z}$ is a 1-form. All components of the metric
and gauge potential are independent of $\psi $, and $u$, $\lambda $, $\sigma 
$, $\phi $ must satisfy
\begin{equation}
\partial_x u =-{\frac{1}{\sqrt{2}\ell }}(\lambda ^{-2}+\sigma ^{-2})
\label{cc1x}
\end{equation}
and 
\begin{eqnarray}
\partial _{z}\partial _{\bar{z}} (\lambda^{-2}-\sigma^{-2})
+ e^{2u} \bigg( \partial_x^2 (\lambda^{-2}-\sigma^{-2})
+3 (\lambda^{-2}-\sigma^{-2}) \partial_x^2 u
\nn
+3 (\lambda^{-2}-\sigma^{-2}) (\partial_x u)^2 +3 \partial_x u \partial_x (\lambda^{-2}-\sigma^{-2})
+{1 \over 2} \ell^{-2} (\lambda^{-2}-\sigma^{-2})^3 \bigg)=0
\label{cc2x}
\end{eqnarray}
and 
\begin{equation}
\partial _{z}\partial _{\bar{z}} u + e^{2u} \bigg( \partial_x^2 u + {1 \over 2} (\partial_x u)^2
+{3 \over 4} \ell^{-2}(\lambda^{-2}-\sigma^{-2})^2 \bigg)=0
\label{cc3x}
\end{equation}
and 
\begin{eqnarray}
d\phi &=&-{\frac{i}{(\lambda \sigma )^{2}}}(\partial _{z}\log \frac{\lambda 
}{\sigma })dx\wedge dz+{\frac{i}{(\lambda \sigma )^{2}}}(\partial _{\bar{z}}
\log \frac{\lambda }{\sigma })dx\wedge d\bar{z}  \notag \\
&&+{\frac{i e^{2u}}{(\lambda \sigma )^{2}}}\left( 2\partial _{x}\log 
\frac{\sigma }{\lambda }+\sqrt{2}\ell ^{-1}\left( \frac{{\lambda ^{2}}-
{\sigma ^{2}}}{(\lambda \sigma )^{2}}\right) \right) dz\wedge d\bar{z}\ .  \label{cc4x}
\end{eqnarray}

In order to recover the anti-self-dual solutions found in \cite{firstpaper} corresponding
to this class of solutions, note first that $F$ is anti-self-dual 
{\footnote{Positive orientation is fixed by $\epsilon_{1 \bo 2 \bt}=-1$} if and only if
\be
\partial_z \lambda =0, \qquad \partial_x \lambda = -{1 \over 2 \sqrt{2} \ell} \lambda^{-1} \ .
\ee
In this case, ({\ref{cc2x}}) is implied by ({\ref{cc1x}}) and ({\ref{cc3x}}). In order to simplify
({\ref{cc3x}}), it is useful to set
\be
x= {1 \over y}, \qquad  u = -2 \log y + {w \over 2} \ .
\ee
Then ({\ref{cc3x}}) is equivalent to
\be
\partial_z \partial_{\bar{z}} w + \partial_y \partial_y e^w =0
\ee
i.e. $w$ is a solution of the $SU(\infty)$ toda equation. Furthermore, if one defines $V$ via
\be
\sigma^{-2} = -\sqrt{2} \ell^{-1} y V
\ee
then ({\ref{cc1x}}) relates $V$ to $w$ by
\be
-2 \ell^{-2} V = y \partial_y w -2
\ee
and the metric can be written as
\be
ds^2 = {1 \over y^2} \bigg( V^{-1} (d \psi+\phi)^2+V (dy^2+ 4e^w dz d {\bar{z}}) \bigg)
\ee
where ({\ref{cc2}}) implies that
\be
d \phi = i \partial_z V dy \wedge dz -i \partial_{\bar{z}} V dy \wedge d {\bar{z}}
+2i \partial_y (e^w V) dz \wedge d {\bar{z}} \ .
\ee
This solution corresponds to one of the anti-self-dual solutions found in \cite{firstpaper}.

We remark that the solutions for which $c_{1}=-{\frac{1}{4}},c_{2}=
-{\frac{i}{2\ell }}$ are found using an essentially identical analysis as given
above. The metric and gauge potential are given by ({\ref{fmm1}}), on making
the replacement $\sigma \rightarrow i\sigma $, and taking $ds^{2}\rightarrow
-ds^{2}$ (to restore the metric signature to $(+,+,+,+)$). The functions $u$, 
$\lambda $, $\sigma $ and the 1-form $\phi $ then satisfy the conditions 
({\ref{cc1x}}), ({\ref{cc2x}}), ({\ref{cc3x}}) and ({\ref{cc4x}}), again
making the replacement $\sigma \rightarrow i\sigma $.

\section{Solutions with $c_1=-{\frac{ 1 }{4}}, c_2=-{\frac{1 }{2 \ell}}$}

For these solutions, one can make a gauge transformation of the $U(1)$ connection $A$, 
which acts on spinors as $\epsilon \rightarrow e^{h}\epsilon $, where $h$ is a real function, and take, without loss of generality

\begin{equation}
\epsilon =1+\sigma e_{1}.
\end{equation}
On evaluating the linear system, one finds the following geometric
constraints 
\begin{eqnarray}
\partial _{1}\sigma -\sigma \Omega _{1,2\bar{2}} &=&0,  \notag \\
\partial _{2}\sigma +\sigma \Omega _{2,1\bar{1}} &=&0,  \notag \\
\Omega _{1,2\bar{1}}=\Omega _{\bar{1},12}=\Omega _{2,12}=\Omega _{2,2\bar{1}} 
&=&0,  \notag \\
\Omega _{2,1\bar{1}}+\Omega _{2,2\bar{2}}-\Omega _{1,12} &=&0,  \notag \\
-\Omega _{2,1\bar{1}}+\Omega _{\bar{1},2\bar{1}}+\Omega _{2,2\bar{2}} &=&0, 
\notag \\
\Omega _{1,1\bar{1}}-\Omega _{1,2\bar{2}}+\Omega _{\bar{2},2\bar{1}} &=&0, 
\notag \\
\Omega _{\bar{2},12}+\Omega _{1,1\bar{1}}+\Omega _{1,2\bar{2}}-\sqrt{2}\ell
^{-1}\sigma &=&0,  \notag \\
\Omega _{1,2\bar{2}}-\Omega _{\bar{1},2\bar{2}} &=&{\frac{1}{\sqrt{2}}}\ell
^{-1}(\sigma -\sigma ^{-1}),  \notag \\
\Omega _{1,1\bar{1}}-\Omega _{\bar{1},1\bar{1}} &=&{\frac{1}{\sqrt{2}}}\ell
^{-1}(\sigma +\sigma ^{-1}),  \label{geox2y}
\end{eqnarray}
together with the following constraints on the gauge potential and field
strength 
\begin{equation}
\ell ^{-1}A_{1}={\frac{1}{2}}(\Omega _{1,1\bar{1}}+\Omega _{1,2\bar{2}}),
\qquad \ell ^{-1}A_{2}=-{\frac{1}{2}}(\Omega _{2,1\bar{1}}
+\Omega _{2,2\bar{2}})  \label{gpotx1y}
\end{equation}

and 
\begin{eqnarray}
F_{1\bar{1}} &=&\ell ^{-1}-{\frac{1}{\sqrt{2}}}\left( (\sigma +\sigma
^{-1})\Omega _{1,1\bar{1}}-(\sigma -\sigma ^{-1})\Omega _{1,2\bar{2}}\right)
,  \label{gfs2y} \\
F_{2\bar{2}} &=&{\frac{1}{\sqrt{2}}}\left( (\sigma -\sigma ^{-1})
\Omega _{1,1\bar{1}}-(\sigma +\sigma ^{-1})\Omega _{1,2\bar{2}}\right) ,  \notag \\
F_{12} &=&-{\frac{1}{\sqrt{2}\sigma }}\Omega _{1,12},  \notag \\
F_{\bar{1}2} &=&{\frac{1}{\sqrt{2}}}\sigma \Omega _{\bar{1},2\bar{1}}. 
\notag
\end{eqnarray}
We remark that ({\ref{gpotx1y}}) relates the $U(1)$ connection $A$ to
the spin connection $\Omega$, and
leads to a partial cancellation of these two connections in the Killing
spinor equation. This is similar to what happens in twisted field theories.

To proceed, it is useful to define 
\begin{equation}
V=\mathbf{e}^{1}+\mathbf{e}^{\bar{1}}
\end{equation}
and denote the vector field dual to $V$ by $V={\frac{\partial }{\partial
\psi }}$. Then ({\ref{geox2y}}) implies that 
\begin{equation}
{\frac{\partial \sigma }{\partial \psi }}={\frac{1}{\sqrt{2}\ell }}(\sigma
^{2}-1).
\end{equation}
There are therefore two subcases, corresponding to $\sigma ^{2}\neq 1$ and $\sigma ^{2}=1$.

\subsection{Solutions with $\protect\sigma^2=1$}

The solutions with $\sigma =-1$ are gauge-equivalent to those with $\sigma
=1 $, so it suffices to take $\sigma =1$. In this case, one has 
\begin{equation}
i(\mathbf{e}^{1}-\mathbf{e}^{\bar{1}})=
\sqrt{2}e^{-\sqrt{2}\ell ^{-1}\psi }{\hat{\mathbf{e}}}^{1},
\qquad \mathbf{e}^{2}=
e^{-\sqrt{2}\ell ^{-1}\psi }{\hat{\mathbf{e}}}^{2}
\end{equation}
where 
\begin{equation}
\mathcal{L}_{V}{\hat{\mathbf{e}}}^{1}=\mathcal{L}_{V}{\hat{\mathbf{e}}}^{2}=0 \ .
\end{equation}
On setting 
\begin{equation}
\mathbf{e}^{1}+\mathbf{e}^{\bar{1}}=2d\psi +\Phi ,
\end{equation}
one then finds that 
\begin{equation}
d{\hat{\mathbf{e}}}^{1}=\Psi \wedge {\hat{\mathbf{e}}}^{1},\qquad d{\hat{\mathbf{e}}}^{2}
=\Psi \wedge {\hat{\mathbf{e}}}^{2}
\end{equation}
where 
\begin{eqnarray}
\Psi &=&-{\frac{1}{\sqrt{2}}}\ell ^{-1}\Phi -\sqrt{2}ie^{-\sqrt{2}\ell
^{-1}\psi }(\Omega _{1,1\bar{1}}+{\frac{1}{\sqrt{2}}}\ell ^{-1}){\hat{\mathbf{e}}}^{1}\newline
\notag \\
&&-e^{-\sqrt{2}\ell ^{-1}\psi }\left( \Omega _{2,2\bar{2}}{\hat{\mathbf{e}}}^{2}
-\Omega _{\bar{2},2\bar{2}}{\hat{\mathbf{e}}}^{\bar{2}}\right)
\end{eqnarray}
satisfies 
\begin{equation}
\mathcal{L}_{V}\Psi =0.
\end{equation}
It follows that one can introduce further local co-ordinates $x,z$, where $x$
is real and $z$ is complex, and a real function $H=H(x,z,{\bar{z}})$ such
that 
\begin{equation}
{\hat{\mathbf{e}}}^{1}=Hdx,\quad {\hat{\mathbf{e}}}^{2}=Hdz
\end{equation}
such that 
\begin{equation}
\Psi =d\log H \ .
\end{equation}
Furthermore, the geometric conditions imply 
\begin{equation}
\mathcal{L}_{V}\Phi =\sqrt{2}\ell ^{-1}\Phi +2d\log H
\end{equation}
and 
\begin{equation}
{\tilde{d}}\Phi =-d\log H\wedge \Phi
\end{equation}
where ${\tilde{d}}$ denotes the exterior derivative restricted to
hypersurfaces of constant $\psi $. It then follows that 
\begin{equation}
\Phi =-\sqrt{2}\ell H^{-1}dH+e^{\sqrt{2}\ell ^{-1}\psi }H^{-1}d\chi
\end{equation}
where $\chi $ is a function of $x,z,\bar{z}$. The gauge potential is then
given by 
\begin{equation}
\ell ^{-1}A={\frac{1}{\sqrt{2}}}\ell ^{-1}d\psi -{\frac{1}{2}}d\log H+
{\frac{1}{\sqrt{2}}}\ell ^{-1}e^{\sqrt{2}\ell ^{-1}\psi }H^{-1}d\chi .
\end{equation}
The solution can then be simplified further by making the co-ordinate
transformation 
\begin{equation}
\psi =\psi ^{\prime }+{\frac{1}{\sqrt{2}}}\ell \log H
\end{equation}
and dropping the prime on $\psi ^{\prime }$ to obtain 
\begin{equation}
ds^{2}={\frac{1}{2}}\left( 2d\psi +e^{\sqrt{2}\ell ^{-1}\psi }d\chi \right)
^{2}+e^{-2\sqrt{2}\ell ^{-1}\psi }ds^{2}(\mathbb{R}^{3})
\end{equation}
with 
\begin{equation}
F=d\left( {\frac{1}{\sqrt{2}}}e^{\sqrt{2}\ell ^{-1}\psi }d\chi \right) .
\end{equation}
Imposing the gauge field equations $d\star F=0$ implies that 
\begin{equation}
\Box _{3}\chi =0
\end{equation}
where $\Box _{3}$ denotes the Laplacian on $\mathbb{R}^{3}$. This is the
Euclidean analogue of the Kastor-Traschen solution \cite{firstpaper, KS}.

\subsection{Solutions with $\protect\sigma ^{2}\neq 1$}

For these solutions, 
\begin{equation}
\sigma =-\tanh \left( {\frac{\psi }{\sqrt{2}\ell }}+h\right)
\end{equation}
where $h$ is a function such that ${\frac{\partial h}{\partial \psi }}=0$.
By making a re-definition of $\psi $, one can, without loss of generality,
set $h=0$. Next, observe that ({\ref{geox2y}}) implies that 
\begin{eqnarray}
\mathcal{L}_{V}(i(\mathbf{e}^{1}-\mathbf{e}^{\bar{1}})) &=&-{\frac{1}{\sqrt{2}\ell }}
(\sigma ^{-1}+\sigma )i(\mathbf{e}^{1}-\mathbf{e}^{\bar{1}}),\qquad 
\notag \\
\mathcal{L}_{V}\mathbf{e}^{2} &=&-{\frac{1}{\sqrt{2}\ell }}(\sigma
^{-1}+\sigma )\mathbf{e}^{2}.
\end{eqnarray}
It follows that 
\begin{equation}
i(\mathbf{e}^{1}-\mathbf{e}^{\bar{1}})={\frac{\sqrt{2}\sigma }{1-\sigma ^{2}}}
{\hat{\mathbf{e}}}^{1},\qquad \mathbf{e}^{2}={\frac{\sigma }{1-\sigma ^{2}}}
{\hat{\mathbf{e}}}^{2}
\end{equation}
where ${\hat{\mathbf{e}}}^{1}$ is a real 1-form, and ${\hat{\mathbf{e}}}^{2}$
is a complex 1-form such that 
\begin{equation}
\mathcal{L}_{V}{\hat{\mathbf{e}}}^{1}=\mathcal{L}_{V}{\hat{\mathbf{e}}}^{2}=0.
\end{equation}
It is convenient to define the $\psi $-independent 3-metric on the
3-manifold $GT$ corresponding to the space of orbits of $V$ by 
\begin{equation}
ds_{GT}^{2}=({\hat{\mathbf{e}}}^{1})^{2}+2{\hat{\mathbf{e}}}^{2}{\hat{\mathbf{e}}}^{\bar{2}}\ ,
\end{equation}
then ({\ref{geox2y}}) implies that $GT$ admits a $\psi $-independent real
orthonormal basis $E^{i}$ ($i=1,2,3$) such that 
\begin{equation}
dE^{i}=\ell ^{-1}\star _{3}E^{i}+\mathcal{B}\wedge E^{i}  \label{gtstrucy}
\end{equation}
where $\star _{3}$ is the Hodge dual on $GT$ and 
\begin{eqnarray}
\mathcal{B} &=&{\frac{i}{\sqrt{2}}}{\frac{\sigma ^{2}}{(1-\sigma ^{2})^{2}}}
\left[ \left( \sigma ^{-1}+\sigma \right) \left( \Omega _{1,2\bar{2}}+\Omega
_{\bar{1},2\bar{2}}\right) -(\sigma ^{-1}-\sigma )\left( \Omega _{1,1\bar{1}}
+\Omega _{\bar{1},1\bar{1}}\right) \right] {\hat{\mathbf{e}}}^{1}  \notag \\
&&+{\frac{\sigma ^{2}}{(1-\sigma ^{2})^{2}}}\left[ -(\sigma ^{-1}-\sigma
)\Omega _{2,2\bar{2}}+(\sigma ^{-1}+\sigma )\Omega _{2,1\bar{1}}\right] {\hat{\mathbf{e}}}^{2} 
 \notag \\
&&+{\frac{\sigma ^{2}}{(1-\sigma ^{2})^{2}}}\left[ (\sigma ^{-1}-\sigma)
\Omega _{\bar{2},2\bar{2}}+(\sigma ^{-1}+\sigma )\Omega _{\bar{2},1\bar{1}}
\right] {\hat{\mathbf{e}}}^{\bar{2}}.
\end{eqnarray}
It follows that $GT$ admits a hyper-CR Einstein-Weyl structure \cite{GT98}. Note that
({\ref{gtstrucy}}) implies that 
\begin{equation}
\mathcal{L}_{V}\mathcal{B}=0
\end{equation}
and 
\begin{equation}
d\mathcal{B}=-\ell ^{-1}\star _{3}\mathcal{B}.
\end{equation}
Next, write 
\begin{equation}
\mathbf{e}^{1}+\mathbf{e}^{\bar{1}}=2d\psi +\Phi
\end{equation}
where $\Phi $ is a 1-form on $GT$; note that ({\ref{geox2y}}) implies that 
\begin{equation}
\mathcal{L}_{V}\Phi =2\mathcal{B}+{\frac{1}{\sqrt{2}\ell }}(\sigma +\sigma
^{-1})\Phi \ .
\end{equation}
Hence 
\begin{equation}
\Phi =-{\frac{\ell }{\sqrt{2}}}(\sigma ^{-1}+\sigma )\mathcal{B}+(\sigma
^{-1}-\sigma )\xi
\end{equation}
where $\xi $ is a $\psi $-independent 1-form on $GT$, with 
\begin{eqnarray}
\xi &=&{\frac{i\ell \sigma ^{2}}{2(1-\sigma ^{2})^{2}}}\left[ \left( \sigma
^{-1}-\sigma \right) \left( \Omega _{1,2\bar{2}}+\Omega _{\bar{1},2\bar{2}}
\right) -\left( \sigma ^{-1}+\sigma \right) \left( \Omega _{1,1\bar{1}}
+\Omega _{\bar{1},1\bar{1}}\right) \right] \mathbf{\hat{e}}^{1}  \notag \\
&&+{\frac{\ell \sigma ^{2}}{\sqrt{2}(1-\sigma ^{2})^{2}}}\left[ \left(
\sigma ^{-1}-\sigma \right) \Omega _{2,1\bar{1}}-\left( \sigma +\sigma
^{-1}\right) \Omega _{2,2\bar{2}}\right] \mathbf{\hat{e}}^{2}  \notag \\
&&+{\frac{\ell \sigma ^{2}}{\sqrt{2}(1-\sigma ^{2})^{2}}}\left[ -\left(
\sigma ^{-1}-\sigma \right) \Omega _{\bar{2},1\bar{1}}+\left( \sigma +\sigma
^{-1}\right) \Omega _{\bar{2},2\bar{2}}\right] \mathbf{\hat{e}}^{\bar{2}}.
\end{eqnarray}
The remaining content of ({\ref{geox2y}}) implies that 
\begin{equation}
d\xi +\mathcal{B}\wedge \xi =-\ell ^{-1}\star _{3}\xi .
\end{equation}
The gauge potential is determined by ({\ref{gpotx1y}}) as 
\begin{equation}
\ell ^{-1}A={\frac{1}{2}}d\log \sigma +{\frac{1}{2\sqrt{2}\ell }}(\sigma
^{-1}+\sigma )d\psi -{\frac{1}{4}}(\sigma ^{2}+\sigma ^{-2})\mathcal{B}
+{\frac{1}{2\sqrt{2}\ell }}(\sigma ^{-2}-\sigma ^{2})\xi .
\end{equation}
On taking the exterior derivative of this expression, and using the
geometric constraints described above, one obtains the components of $F$
given in ({\ref{gfs2y}}); moreover, the gauge field equations hold with no
additional constraints.

To summarize, the metric is given by 
\begin{equation}
\label{st1}
ds^{2}={\frac{1}{2}}\left( {\frac{2\sqrt{2}\ell }{\sigma ^{2}-1}}d\sigma -
{\frac{\ell }{\sqrt{2}}}(\sigma ^{-1}+\sigma ) {\cal{B}}+(\sigma ^{-1}-\sigma )\xi
\right) ^{2}+{\frac{\sigma ^{2}}{(1-\sigma ^{2})^{2}}}ds_{GT}^{2}
\end{equation}
where we have changed co-ordinates from $\psi $ to $\sigma $. $GT$ is a
3-manifold, which admits a hyper-CR Einstein-Weyl structure; in particular, there is
a $\sigma $-independent real basis $E^{i}$ such that 
\begin{equation}
dE^{i}=\ell ^{-1}\star _{3}E^{i}+\mathcal{B}\wedge E^{i}  \label{gtd1}
\end{equation}
where $B$ is a $\sigma $-independent 1-form on $GT$ satisfying 
\begin{equation}
d {\cal{B}}=-\ell ^{-1}\star _{3}\mathcal{B}  \label{gtd2}
\end{equation}
and $\xi $ is a $\sigma $-independent 1-form on $GT$ satisfying 
\begin{equation}
d\xi +\mathcal{B}\wedge \xi =-\ell ^{-1}\star _{3}\xi .  \label{gtd3}
\end{equation}
The gauge field strength is given by 
\begin{equation}
F=d\left[ -{\frac{1}{4}}\ell (\sigma ^{2}+\sigma ^{-2})\mathcal{B}
+{\frac{1}{2\sqrt{2}}}(\sigma ^{-2}-\sigma ^{2})\xi \right] \ .  \label{gfst}
\end{equation}
The field strength $F$ is anti-self-dual if and only if $\xi = {\ell \over \sqrt{2}} {\cal{B}}$.

We remark that the solutions for which $c_{1}=-{\frac{i}{4}},c_{2}=
-{\frac{i}{2\ell }}$ are determined using an almost identical analysis to that given
in this section, so we again simply summarize the result. In this case, the
metric is given by 
\begin{equation}
ds^{2}={\frac{1}{2}}\left( {\frac{2\sqrt{2}\ell }{1+\sigma ^{2}}}d\sigma -
{\frac{1}{\sqrt{2}}}\ell (\sigma ^{-1}-\sigma )\mathcal{B}+(\sigma
^{-1}+\sigma )\xi \right) ^{2}+{\frac{\sigma ^{2}}{(1+\sigma ^{2})^{2}}}
ds_{GT}^{2} \ ,
\end{equation}
$GT$ is a 3-manifold, which admits a hyper-CR Einstein-Weyl structure; there is a 
$\sigma $-independent real basis $E^{i}$ satisfying ({\ref{gtd1}}), where 
$\mathcal{B}$ is a $\sigma $-independent 1-form on $GT$ satisfying 
({\ref{gtd2}}), and $\xi $ is a $\sigma $-independent 1-form on $GT$ satisfying 
({\ref{gtd3}}). Furthermore, the gauge field strength is given by ({\ref{gfst}}).

\section{Conclusions}

In this work, we have classified all local forms of supersymmetric solutions to
four-dimensional Einstein-Maxwell Euclidean supergravity, for which the
Maxwell field strength is neither anti-self-dual or self-dual. The backgrounds
we have considered admit a Killing spinor which satisfies a Killing spinor equation, whose
structure was fixed by requiring that the integrability conditions of the Killing spinor
equation should be compatible with the Einstein field equations of an Einstein-Maxwell
Lagrangian. We have therefore extended the earlier classification of \cite{firstpaper},
in which the solutions for which the Maxwell field is anti-self-dual were classified.

The solutions which we have found fall into three classes. The first class of solutions
corresponds to the solution given in equations ({\ref{fmm1}})-({\ref{cc4x}}) of section 3.
This solution is a Euclidean version of the solution originally found in the ``timelike" class 
of the Lorentzian $N=2$, $D=4$ supergravity in \cite{lorentz1, lorentz2}.
In particular, we remark that the equations ({\ref{cc1x}})-({\ref{cc3x}}) which determine the
solution are, under a rescaling of $\lambda, \sigma$, identical to equations (2.9) and (2.10)
of \cite{lorentz2}, with the exception of the sign of the term involving $(\lambda^{-2} - \sigma^{-2})^2$
in ({\ref{cc3x}), which differs from the sign of $B^2$ in equation (2.10) of \cite{lorentz2}.
All of the solutions in section 3 preserve two (real) supersymmetries.
The remaining two classes of solution are obtained in section 4. One of these, derived in
section 4.1,
is the Euclidean version of the Kastor-Traschen solution. The remaining class of solutions 
obtained in section 4.2 does not have an analogous solution in the Lorentzian theory, and
is constructed in ({\ref{st1}})-({\ref{gfst}}) from a 3-dimensional base space which is
a 3-manifold admitting a hyper-CR Einstein-Weyl structure. 

All of the solutions in section 4 automatically preserve four (real) supersymmetries;
this can be seen by observing that if the constants $c_1, c_2$ appearing in the Killing spinor equation
({\ref{gkse2b}}) are both real, and $\epsilon$ is a Killing spinor, then so is $C* \epsilon$, where
$C$ is an appropriately constructed charge conjugation operator. If, however, $c_1$, $c_2$ are
purely imaginary and $\epsilon$ solves ({\ref{gkse2b}}), then $C* \gamma_5 \epsilon$ also solves
({\ref{gkse2b}}). This type of automatic supersymmetry enhancement does not occur for the
cases described in section 3, where one of $c_1, c_2$ is real, and the other imaginary.
The analysis of solutions of Euclidean four-dimensional supergravity with enhanced supersymmetry
is currently work in progress.

\bigskip

\textbf{Acknowledgements}. MD, JG and PT thank the American University of
Beirut for hospitality when some of this work was carried out. The work of
WS is supported in part by the National Science Foundation under grant
number PHY-0903134. JG is supported by the EPSRC grant EP/F069774/1

\end{document}